# PROTOTYPING BIO-NANOROBOTS USING MOLECULAR DYNAMIC SIMULATION

*Mustapha Hamdi [1], Gaurav Sharma [2], Antoine Ferreira [1], Constantinos Mavroidis [2]*

[1]Laboratoire Vision et Robotique, Université d'Orléans, 18000, Bourges, France.
[2]Computational BioNano Robotics Laboratory, Northeastern University, Boston, MA-02120, USA
antoine.ferreira@ensi-bourges.fr , mavro@coe.neu.edu

**ABSTRACT**

This paper presents a molecular mechanics study using a molecular dynamics software (NAMD) coupled to virtual reality techniques for intuitive bio-nanorobotic prototyping. Using molecular dynamic simulations the operator can design, characterize and prototype the behavior of bio-nanorobotic components and structures. The main novelty of the proposed simulations is based on the characterization of stiffness performances of passive joints-based deca-alanine protein molecule and active joints-based viral protein motor (VPL) in their native environment. Their use as elementary bio-nanorobotic components are also simulated and the results discussed.

## 1. INTRODUCTION

Bio-nanorobotics is an emerging area of scientific and technological opportunity. It is a new and rapidly growing interdisciplinary field addressing the assembly, construction and utilization of biomolecular devices based on nanoscale principles and/or dimensions. Research and product development at the interface of physical sciences and biology as applied to this area require multi-skilled teams and often novel technical approaches for material synthesis, characterization and applications. In this way proteins and DNA could act as motors, mechanical joints, transmission elements, or sensors. If all of these different components were assembled together they can form bio-nanorobots with multiple degrees-of-freedom that are able to apply forces and manipulate objects in the nanoscale world. These motors, which are called biomolecular motors, have attracted a great deal of attention recently because they have high efficiency, they could be self-replicating, and hence cheaper in mass usage, and they are readily available in nature. A number of *enzymes* such as kinesin [1], RNA polymerase [2], myosin [3], dynein [4], adenosine triphosphate (ATP) synthase [5], viral protein linear (VPL) motor [6] and DNA can function as nanoscale linear, oscillatory or rotary biological motors. In addition, there are compliance devices such as spring-like proteins called fibronectin [7] and vorticellids [8], as well as synthetic contractile plant polymers [9] which can act as compliant joints in biomolecular robotic systems. The idea is to use biomolecular motors as the actuators of such bio-nanorobots, where the structural elements are carbon nanotubes, while the joints are formed by appropriately designed DNA elements. Figure 1a and 1b show a schematic of VPL motor supporting a moving platform. Precise manipulation could help scientists better understand the principles of molecular motors, at the origin of the bio-nanorobot actuation. Going beyond a knowledge of the forces derived from these nanomanipulation experiments requires a contribution from molecular dynamic simulations to be able to understand the bio-nanomechanics of proteins and develop dynamic and kinematic models to study their performances. The ability to visualize the atom-to-atom interaction in real-time and see the results in a fully immersive 3-D environment is an additional feature of such simulations [14-18]. In this work, we consider a mechanical molecular study of unfolding proteins acting as passive or active joints in bio-nanorobotic systems. We therefore decided to begin our investigations by simulating the forces involved under various external mechanical stress (stretching, contraction, shearing, bending) to predict the type of force spectra, reversibility, degrees of freedom and irreversible work that may be expected from single-molecule protein manipulation experiments.

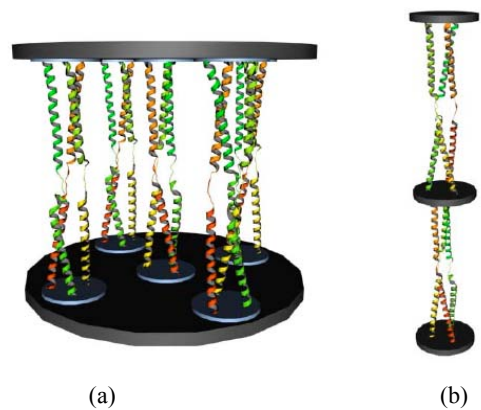

(a)           (b)

**Fig. 1:** Several VPL motors placed in parallel (left) and series (right) to multiply force and displacement respectively.

The paper is organized as follows: Section II presents mechanical molecular simulations of deca-alanine and viral proteins as passive/active links before to make a discussion about the current developments in bio-nanorobotics field in Section 3.

## 2. SIMULATION OF BIO-NANOROBOTIC COMPONENTS

**2.1. Simulation of Elementary Bionanorobotic Module**

Figure 2 shows a schematic of an elementary bionanorobotic module composed of a moving platform supported by the VPL motor and two deca-alanine proteins. The first image in Figure 4a shows the VPL in its initial state and the second image shows it in the final extended state due to pH activation. For detailed information on the working of the VPL motor the reader is encouraged to read reference [6]. Two deca-alanine proteins can be used as passive spring elements to join two platforms and form a single d.o.f parallel platform that is actuated by the upward stretching of the VPL actuator.

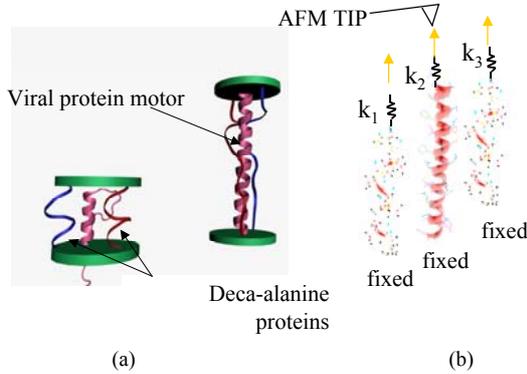

(a)      (b)
**Fig. 2:** One degree of freedom Bio-Nanorobotic module actuated by a viral protein motor and two α-helix deca-alanine protein springs.

Their elastic behavior can be used as a passive control element or as the restitution force that will bring the platform back to its original position. The proposed MD and VR system allows to simulate proteins and theoretically evaluate the forces generated in biochemical reactions, and even to exert external forces to alter the fate of these reactions. We present below the simulated mechanical properties of different passive and active bio-nanorobotic components of Figure 2a when subjected to external forces. The external forces (stretching, shearing or bending) are applied using a force feedback device (shown as a virtual AFM tip in figures) in order to study the mechanical properties of the principal architectural elements [19]. The various deformations that have been studied use an adiabatic energy mapping.

**2.2. α–Helix Deca-Alanine Protein as Passive Joint**

We applied the techniques described above to an exemplary system, helix-coil transition of deca-alanine in vacuum and in solvent. The well-known deca-alanine is an oligopeptide composed of ten alanine residues (Figure 5). In vacuum at room temperature ($T$=300°K), the stable configuration of deca-alanine is a α-helix. Stretching the molecule by an external force can induce its transition to an extended (coiled) form.

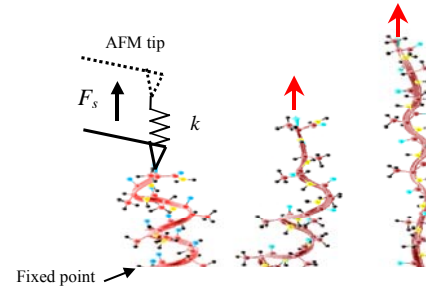

**Fig. 3:** Unfolding of helical deca-alanine. Left, a folded configuration (α-helix) is shown. The six hydrogen bonds that stabilize the helix are shown. Middle and right, extended configurations (coil) when stretching the molecule by an external force (shown as a virtual AFM tip).

This helix-coil transition represents a simple but basic folding system acting as a passive like-spring. In the simulation, we fix one end of the molecule (the main chain nitrogen atom of the first residue) at the origin and constrain the other end (the capping nitrogen atom of the C-terminus) to move only along the $z$ axis, thereby removing the irrelevant degrees of freedom, i.e., removing the overall translation and rotation. For this, a guiding potential [20]

$$u(\mathbf{r};\lambda) = (k/2)[\zeta(\mathbf{r}) - \lambda]^2 \quad (1)$$

is added to control the end-to-end distance ξ which is a function of the 3$N$-dimensional position **r** of the system. The moving guiding potential used in the pulling simulations is represented by a spring which is connected to the C-terminus and pulled with a constant velocity $v$. The parameter λ is varied from 15 to 35 Å with a constant speed $v$ of 0.1 Å/ns. A force constant of $k$=500 pN/Å is used in order to allow end-to-end distance ξ closely follow the constraint center λ. The external work curve is defined as:

$$\langle W \rangle = -kv \int_0^t dt' f(t') \quad (2)$$

Figure 4a-b shows the maximal force and energy variation (before breaking) as a function of stretching distance.

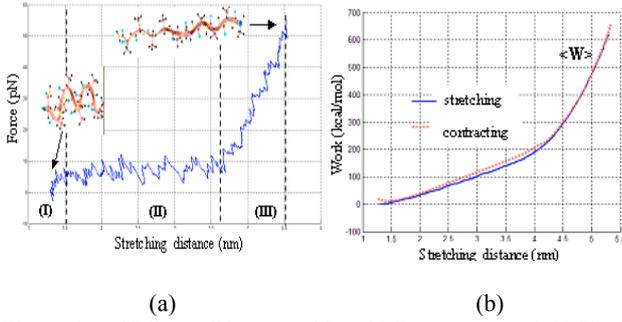

(a)            (b)

**Figure 4:** α-Helix to ribbon stretching. (a) Force curve and (b) Work done by forward pulling (stretching) and backward pulling (contracting) with $v$=0.1 Å/sec. For the forward pulling, the position of the constraint center ξ is varied from 15 to 35Å; for the backward pulling, from 35 to 15 Å.

Depending on the sign of $v$, the external work can be defined for either stretching or contracting motion of the protein. Fig.4a shows two distinct conformational transitions that provoke the conversion of α-helix to an extended form approaching the coil conformation. The force increases rather smoothly to almost 10pN up to a relative length of the protein. Once this transition is completed, the hydrogen bonds start to break and the structure further extends toward the coil conformation (intrinsic elastic regime), as shown in Part II, of the force curve (∼700pN). At the lowest forces, the molecule behaves as a Hookian spring and its extension is proportional to the force applied at the end with a reversible motion. A useful approximation for spring constant $k_{alan}$ is given by the inextensible worm-like chain (*WLC*) model [20]. Then, if the stretching force increases gradually until to reach its stretching limit (∼3600pN) with an irreversible motion in Part III. Fig.4b shows the work done by forward pulling (stretching) and backward pulling at the same speed. As the protein environment has an important influence on its conformation, we studied the influence of the dynamics of water molecules (as a solvent) onto the protein-water interactions. Fig.5 shows the influence of the water-protein interactions on the external force which shows that a greater force (and work) is required for reversible extension of the deca-alanine protein. The water solvent increases considerably the Hookian spring $k_{str}$ of the protein structure. However, as the system contains solvent molecules, the relaxation time increases such that the helix-coil transition can be induced in a reversible manner with a hysteresis behavior.

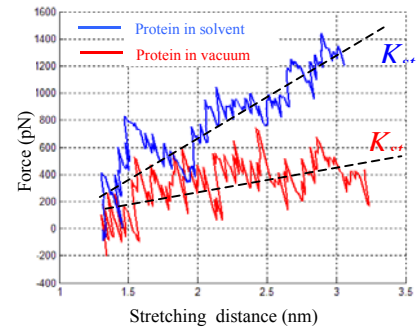

**Figure 5:** Force curve of α-Helix to ribbon done by forward stretching with a reversible motion (Part I and II).

It should be noticed that the use of protein as passive joints suppose that the deca-alanine protein can also be mechanically dependent of temperature versus the work produced by the protein molecule stressed in an unnatural unfolding pathway. The previous results showed that the native structure is not destroyed under normal physiological conditions. In some unnatural conformations, lateral shearing and bending forces applied on a protein molecule has been simulated: fixed-free boundary conditions and fixed-fixed boundary conditions. These tests simulate disturbances of the bio-nanorobotic component under various operating conditions. For these conformations, Fig.6 presents the lateral forces obtained. As shown in Fig.6a, the lateral Hookian spring $k_{shear}$ has high stiff spring value that is able to counteract microenvironment variations and mechanical disturbances of a bio-nanorobotic platform. In this case, in contrast to stretching described above, the force variation is roughly monotonic with different plateau leading to different Hookian spring values. This is explained by a successive rather than a concurrent rupture of the hydrogen bonds joining the strands. Shearing is largely limited to breaking hydrogen bonds as there is little conformational change in the extended peptide backbones. Conversely, Fig.6b shows a linear variation of force behavior when considering pure bending deformation of the protein. It shows that for little length deformation, it requires low constant force. After a threshold value, lateral bending must break the van der Waals interactions of the entire surface, leading to high forces and strong length dependence. Finally, these results indicate that it may be possible to obtain uncoupled mechanical spring behavior of the protein: stiff in lateral and compliant in Furthermore, under normal operating conditions, deca-alanine protein shows a reversibility of displacement-force characteristics which allows its use as spring-like joint in bio-nanorobotic platforms. longitudinal directions.

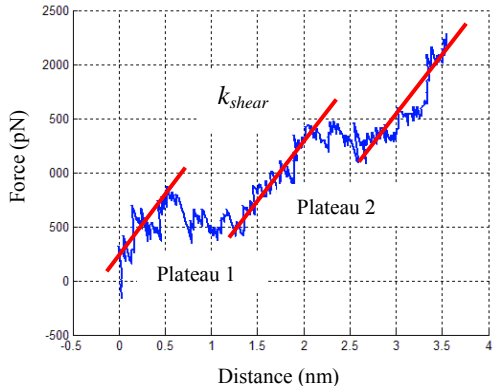

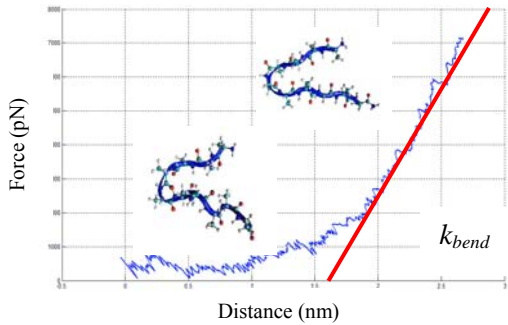

**Fig. 6:** Force curve of α-Helix deca-alannine protein when perturbed by a (a) lateral shearing force $F_{shea}$ and (b) bending force $F_{bend}$.

### 2.3. Viral protein linear (VPL) motor as active joint

In this part, we study the molecular properties of viral proteins to change their 3D conformation depending on the pH level of environment. Thus, a new linear biomolecular actuator type called Viral Protein Linear (VPL) motor, developed originally in [21], is characterized as an active linear joint. The structure is like a hairpin composed of three coils, having one C terminal (carboxy-end) and undergoes a conformational change induced by mildly acidic conditions (i.e., pH around 5). With the change in pH, the N-terminals pop out of the inner side and the peptide acquires a straightened position. The VPL motor uses the Influenza virus protein Hemagglutinin (HA). Two known states are established, i.e., the native state and the fusogenic state of the 36-residue peptide of HA. The structural data of these states are obtained from the Protein Data Bank (PDB) [22]. The representative "open" structure can be generated by forcing the structure away from the native conformation (Fig. 8a) to open state (conformation) (Fig. 8b) with constrained high-temperature molecular dynamics. Both the forward (closed to open) and the reverse (open to close) transformations are carried out for two end-point structures of 9Å. This peptide is able to perform repeatable motion controlled by variation of pH.

In order to study the mechanical limits of the reference end-point states with a reversible motion, simulations of the mechanical stretching through a guiding potential of Eq.2 has been investigated.

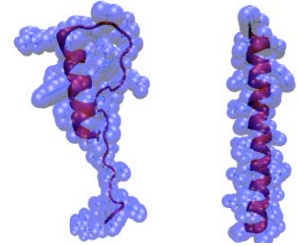

(a)    (b)

**Fig.7:** (a) *loop 36* protein in the native state; (b) open-state conformation of the VPL motor shown with van der Waals surface distribution.

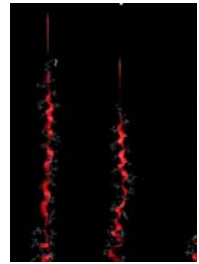 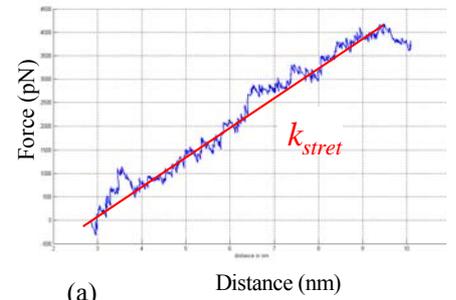

(a)

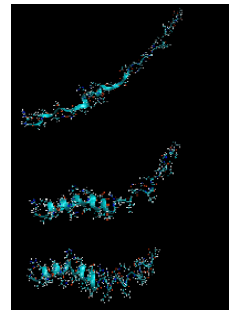 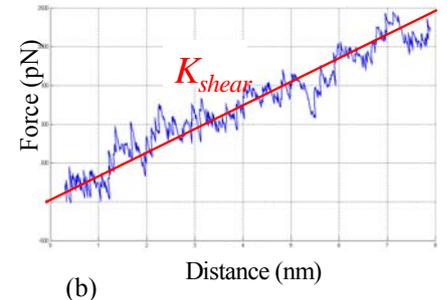

(b)

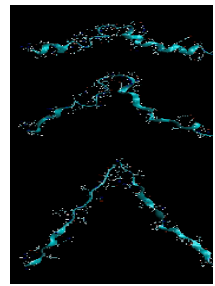 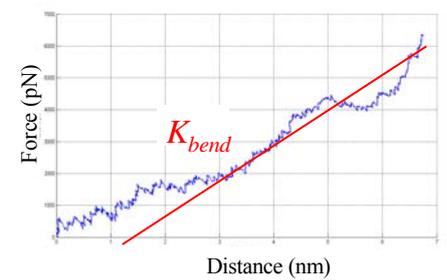

(c)

**Figure 8:** VPL of force-versus-distance characteristics: (a) stretching, (b) shearing and (c) bending for *v*=1Å/ns, *k*=500pN/Å and T=300°K.

Here, we stretch the viral protein in an irreversible manner and examine the resulting distribution of force.

The spring constant is settled at $k$=500pN/Å and T=300°K. The molecule is stretched by changing the parameter λ from 15 to 40 Å with a constant speed. Figure 8a gives the results of displacement and force stretching ($K_{stret}$= 625pN/nm) of the VPL motor. Fig.8b-c shows the shearing ($K_{shear}$= 312pN/nm) and bending results ($K_{bend}$= 714pN/nm) when $v$=1Å/ns, $k$=500pN/Å and T=300°K. A more realistic environment of the VPL requires the inclusion of the effect of pH on the protein. In order to take into account the real-time effect of pH on the ionic stability of the protein, a current model will be considered in a next phase.

### 3. SIMULATION OF A BIO-NANOROBOTIC PARALLEL PLATFORM

In order to prototype mechanical designs of bio-robotic structures, we simulated the parallel nanorobotic platform of Fig.9 of four passive parallel links using serially-linked deca-alanine proteins connected to a graphite platform at both ends. Each link is composed of of three deca-alanine proteins serially linked. The attachment of the different components are 4 carbon atoms (proteins/graphite interface) and 3 hydrogen atoms (protein/protein interface). The dimensions for the ractangular graphite sheet are 14.8nm by 8.5nm and the total length of the three proteins is 13.1nm.

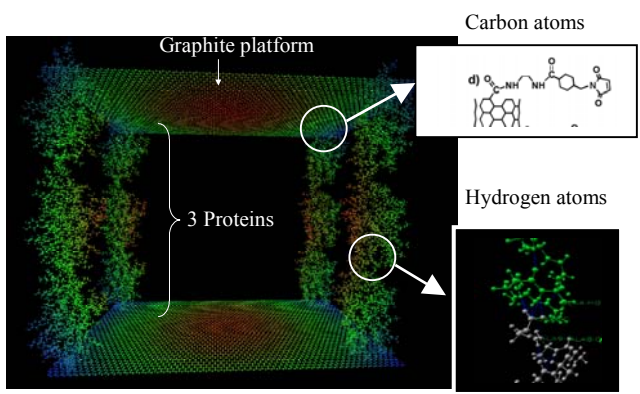

**Fig.9:** Parallel nanorobotic platform composed of two graphite platforms and four serially-linked passive linear links. The four parallel links are composed of three serial deca-alanine proteins.

We applied stretching forces at the upper graphite platform. Simulations showed that the parallel connection of serially-linked deca-alanine proteins permits to augment considerably the resulting spring-force and to decrease the overall displacement. As example, Fig.10 shows the stretching force curve of two parallel links composed of four serial spring-like proteins (deca-alanine) for $v$=1Å/ns, $k$=500pN/Å and T=300°K. The spring force is considerably increased symmetrically for both parallel links until to reach its point of rupture. As expected during stretching (serial/parallel structure) the generated forces in parallel configuration is multiplied by 2 while in serial configuration, the displacement is multiplied roughly by 4. The elongation of the serial configuration is mainly dominated by unfolding the α-helices of the different deca-alanine proteins. However, the rupture point is not due to the rupture of the protein but simply by the rupture of the attachment between the proteins (1,1') and (2,2') as shown in Fig.10. We can see in Fig.11 the linear behavior of the stretching force curve at each protein/protein joint (joints 1, 2, 3 and 4). It should be noted in the Fig.10 that the force behavior of each parallel link is quite similar. As a result of our simulations (stretching, bending and shearing) the different stiffnesses differ strongly depending of the external disturbances applied to the nanorobotic platform:

- stretching stiffness is strongly influenced by the native environment: $(k_{stret})_{air} \cong 400$ pN/nm and $(k_{stret})_{water} \cong 70$ pN/nm
- lateral shearing ($k_{shear} \cong 700$ pN/nm) is much more stiffer than lateral bending ($k_{shear} \cong 30$ pN/nm) which ensures a good mechanical stability of the platform against external disturbances.

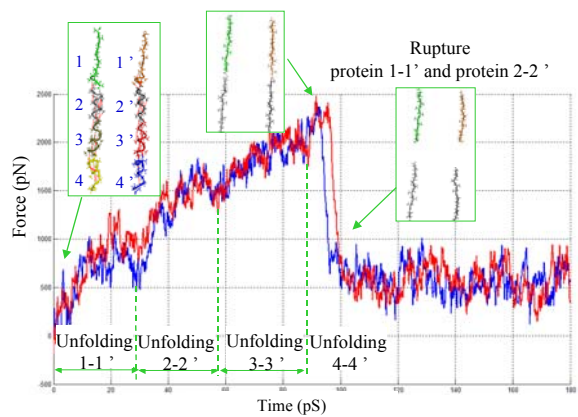

**Fig.10:** Stretching force curve of two parallel links composed of four serial deca-alanine proteins acting as elastic and reversible springs.

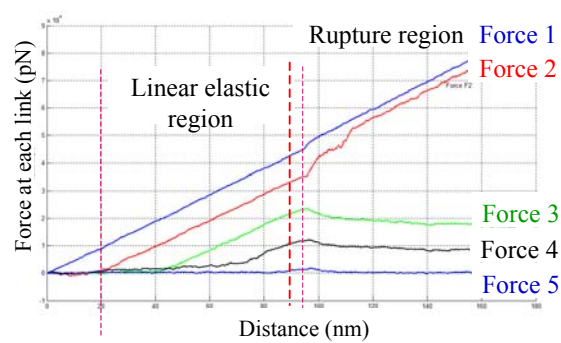

**Fig.11:** Force at each protein link of a four serial deca-alanine proteins acting as elastic and reversible springs.

In order to reduce the number of components, the serial configuration of deca-alanine proteins have been replaced by ROP proteins (parallel link). The repressor of primer protein (ROP) is a small, dimeric molecule consisting of two identical chains of 63 amino acids. The two monomers pack together as a fully antiparallel four helix-bundle. The bend region of Rop has attracted considerable interest as a parallel molecular spring due to its stability and elasticity properties [23]. In the simulation, we constrain both ends of two $\alpha$-helices of the molecule to move only along the *z* axis for stretching simulations and fix the short turn (Fig.12). Compared to the elastic behavior of deca-alanine proteins (see Fig.4a), we can see the same reversible elastic behavior with greater forces due to its parallel structure. Furthermore, each passive link behaves as a perfect reversible linear joint when it is relaxed from its stretched state. Due to its native parallel structure, we need only four ROP proteins connected to the graphite platform.

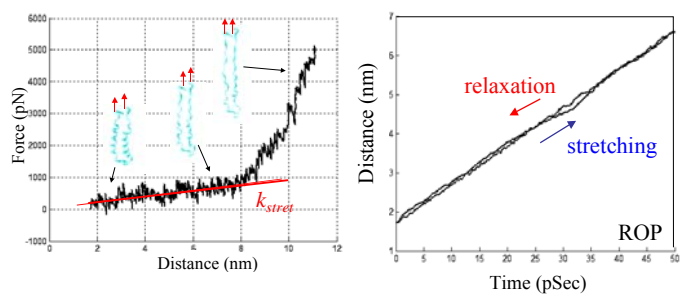

**FIG.12:** Double α-helix to ribbon stretching. (a) Force curve and (b) hysteresis curve done by forward stretching (with *v*=0.1 Å/sec) and relaxation of the structure when the stretching force is zero.

## VI. CONCLUSION

This paper presents a molecular mechanics study carried out with a molecular dynamics computation. It allows simulating the force, position and energy feedback of spring-like and actuator-like proteins for design evaluation of future bio-nanorobots. The preliminary mechanical force results given in this study corroborates the force results when stretching α-helix, β-ribbon proteins and double-helical DNA molecules. Further refinement will be carried out in order to integrate the VPL motor into the parallel platform for active actuation.

REFERENCES

[1] S.M. Block, "Kinesin what gives?", *Cell*, 93, pp.5-8 (1998).
[2] M.D. Wang, M.J. Schnitzer, H.Yin, R. Landrick, J. Gelles, S.M. Block, "Force and velocity measurement for single molecules of RNA polymerase", *Science*, 282, pp.902-907 (1998).
[3] K. Kitamura, M. Tokunaga, A.H. Iwane, Y. Yanagida, "A single myosin head moves along an actin filament with rectangular steps of 5.3nm", *Nature*, 397, PP.129 (1999).
[4] C. Shingyoji, H. Higuchi, M. Yoshimura, E. Katayama, T. Yanagida, "Dynein arms are oscillating force generators", *Nature*, 393, pp.711-714.
[5] C.D. Montegano, G.D. Bachand, "Constructing nanomechanical devices powered by biomolecular motors", *Nanotechnology*, 10, pp.225-331 (1999).
[6] Dubey A., Mavroidis C., Thornton A., Nikitczuk K.P., Yarmush M.L., *Viral Protein Linear (VPL) Nano-Actuators, IEEE Int. Conference on Nanotechnology,* San Francisco, August 12-14, 2003.
[7] H.P. Erickson, "Streching single protein molecules: Titin os a wired spring", *Science*, 276, pp.1090-1092.
[8] L. Mahadevan, P. Matsudaira, "Mobility powered by supramolecular springs and ratchets", *Science*, 288, pp.95-99 (2000)
[9] M. Knoblauch, G.A. Naull, T. Muller, D. Prufer, I. Schneider-Huther, et al. "ATP-independent contractile proteins from plants", *Nature Materials*, 2, pp.600-603 (2003).
[10] T.Yamamoto, O.Kurosawa, H.Kabata, N. Shimamoto, M. Washizu, Molecular surgery of DNA Based on Electrostatic micromanipulation*,* IEEE *Trans. on Industry Applications*, Vol.36, No.4, pp.1010-1017 (2000).
[11] S.Thalhammaer, R. Stark, S.Muller, J.Winberg, W.M., heckl, "The Atomic Force Microscope as a New Microdissecting Tool for the Generation of Genetic Probes*", Journal of Structural Biology*, 119, pp.232-237 (1997).
[14] H. Haase, J. Strassner, and F. Dai, "VR techniques for the investigation of molecule data", *Computers & Graphics*, Special Issue on Virtual Reality, 20(2), pp. 207-217 (1996). Elsevier Science Ltd.
[15] R. C. Drees, J. Pleiss, D. Roller, and R. D. Schmid, "Highly Immersive Molecular Modeling (HIMM): an architecture for the integration of molecular modeling and virtual reality", in *Computer Science and Biology*, Proceedings of the German Conference on Bioinformatics, (Leipzig, Germany), pp. 190-192, Sep-Oct 1996.
[16] *Collaborative Visualization and Simulation Environment (COVISE),* http://www.hlrs.de/organization/vis/covise/ , last updated Mar 30, 2001
[17] W. Humphrey, A. Dalke, K. Schulten, *"VMD:Visual Molecular Dynamics", Journal of Molecular Graphics*, vol.14, pp.33-38, 1996.
[18] Ferreira A., Sharma G., Mavroidis D., ''New trends in Bio-Nanorobotics using Virtual Reality technologies'' *IEEE International Conference on Robotics and Biomimetics*, June30-July3, Hong-Kong (China), 2005, pp.89-94.
[19] M. Hamdi, G. Sharma, A. Ferreira, C. Mavroidis ''Molecular Mechanics Study on Bionanorobotic Components using Force Feedback'' *IEEE International Conference on Robotics and Biomimetics*, June30-July3, Hong-Kong (China), 2005, pp.105-110.
[20] S. Park, F.Khalili-Araghi, E.Tajkhorshid, K.Schulten, "Free energy calculation from steered molecular dynamics simulations using Jarzynski's equality", *Journal of Chemical Physics*, vol.119, n°6, pp.3559-3566, 2003.
[21] A.Dubey, G.Sharma, C.Mavroidis, S.M.Tomassone, K.Nikitckuk, L.Yarmush, "Computational Studies of Viral Protein Nano-Actuators*", Jour. of Comp. and Theoretical Nanoscience*, vol.1, n°.1, pp.1-11, 2004.
[22] H.M. Berman, J.Westbrook, Z.feng, G.Gilliland, T.N. Bhat, H.Weissig, I.N. Shindyalov, and P.E. Bourne, "The Protein Data Bank", *Nucleic Acid Research,* 28, pp.235-242, 2000.
[23] H.P. Kresse, M.Czuayko, G.Nyakatura, G.Vriand, C. Sander, H.Bloecker, "Four-helix bundle topology re-engineered: monomeric Rop protein variants with different loop arrangements", Protein Engineering, Vol.14, N°11, pp.897-901, 2001.